# Diagnosis of COVID-19 Using Machine Learning and Deep Learning: A Review

M. Rubaiyat Hossain Mondal[1,*], Subrato Bharati[1] and Prajoy Podder[1]

[1]Institute of ICT, Bangladesh University of Engineering and Technology, Dhaka-1205, Bangladesh

**Abstract:** *Background*: This paper provides a systematic review of the application of Artificial Intelligence (AI) in the form of Machine Learning (ML) and Deep Learning (DL) techniques in fighting against the effects of novel coronavirus disease (COVID-19).

*Objective & Methods*: The objective is to perform a scoping review on AI for COVID-19 using preferred reporting items of systematic reviews and meta-analysis (PRISMA) guidelines. A literature search was performed for relevant studies published from 1 January 2020 till 27 March 2021. Out of 4050 research papers available in reputed publishers, a full-text review of 440 articles was done based on the keywords of AI, COVID-19, ML, forecasting, DL, X-ray, and Computed Tomography (CT). Finally, 52 articles were included in the result synthesis of this paper. As part of the review, different ML regression methods were reviewed first in predicting the number of confirmed and death cases. Secondly, a comprehensive survey was carried out on the use of ML in classifying COVID-19 patients. Thirdly, different datasets on medical imaging were compared in terms of the number of images, number of positive samples and number of classes in the datasets. The different stages of the diagnosis, including preprocessing, segmentation and feature extraction were also reviewed. Fourthly, the performance results of different research papers were compared to evaluate the effectiveness of DL methods on different datasets.

*Results*: Results show that residual neural network (ResNet-18) and densely connected convolutional network (DenseNet 169) exhibit excellent classification accuracy for X-ray images, while DenseNet-201 has the maximum accuracy in classifying CT scan images. This indicates that ML and DL are useful tools in assisting researchers and medical professionals in predicting, screening and detecting COVID-19.

*Conclusion*: Finally, this review highlights the existing challenges, including regulations, noisy data, data privacy, and the lack of reliable large datasets, then provides future research directions in applying AI in managing COVID-19.

## 1. INTRODUCTION

From December 2019, a novel coronavirus named as severe acute respiratory syndrome coronavirus 2 (SARs-CoV-2) has caused a fatal disease called coronavirus disease (COVID-19) [1-3]. Although COVID-19 started in Wuhan city, China, the whole world is currently heavily suffering from the disease [4]. SARs-CoV-2 has caused more harmful effects than other previous coronaviruses, namely severe acute respiratory syndrome coronavirus (SARS-CoV), which killed 774 people, and Middle East respiratory syndrome (MERS-CoV) killing 858 people [5-7]. Table **1** shows a comparison of COVID-19 and other pandemics rec-orded in human history. Because of the huge number of confirmed cases in COVID-19 [8], the health systems in many countries are becoming overwhelmed, and the demand for intensive care unit (ICU) is increasing beyond capacity.

Due to the severe transmission capacity of the virus and its potential harm, it is important to predict the COVID-19 pandemic. In particular, the prediction of the disease enables a country to respond appropriately for the near future. However, the prediction of this disease has several challenges. Some of these challenges are the tracking of the infected people, no definite treatment option, increased fatality chances for the elderly or people with other serious diseases, 2-week long incubation period, and lack of accuracy in the available datasets, period, and strictness of the lockdown. Hence, the accurate prediction of COVID-19 is necessary.

Recently, there is great interest in finding the appropriate solutions to managing COVID-19. One big challenge is the ever-increasing and changing volume of COVID-19 related data, which creates difficulty in developing optimal solutions. Under these circumstances, Artificial Intelligence (AI) can contribute to managing COVID-19. With its subsets, Machine Learning (ML) and Deep Learning (DL) methods, AI can facilitate mobilisation and save medical, logistic, human resources, and precise time. It can be noted that ML and DL are widely used in a number of medical systems for identifying patterns in data samples. ML and DL can map patterns from large data volumes and improve quickly with new data availability. Moreover, AI can ensure reliability in predicting the spread of the virus, classifying suspected patients, and data-driven diagnosis of COVID-19. Along with big data, AI can be used to model activities. This helps policymakers manage a crisis, plan for treatment strategies, and optimise diagnosis techniques, including medical imaging and image processing. In an AI-based automated COVID-19 detection system, a remote video diagnosis scheme and robotics can play a part in the initial diagnosis. Smart robots can help in patient treatment without involving human. The automated classification of images helps prevent the virus from patients to the radiologists and medical staff. AI can help in the identification of fake data and separate relevant scientific information. AI-based schemes can help monitor and track the COVID-19 patients, which helps in controlling the virus spread. With the DL models' improvement, it will be easy to differentiate between COVID-19 and non-COVID pneumonia. However, AI has not been appropriately employed to serve healthcare systems in the fight against COVID-19. AI specialists can fill the gap between medical treatments and AI-based approaches. In many research studies, DL as a part of AI has been used to develop support systems in diagnosing COVID-19. In this case, both Computed Tomography (CT) and X-ray image samples are considered. Some support systems use pre-trained transfer learning models, and some others use customised DL networks. AI has also been used to diagnose, prognosis, predict and forecast COVID-19. Besides, allied fields such as computer vision, the Internet of Things (IoT), smartphone technology, big data assist AI in developing solutions for fighting COVID-19.

*Address correspondence to this author at the Institute of ICT, Bangladesh University of Engineering and Technology, Dhaka-1205, Bangladesh; Tel: +88 01711014224; E-mail: rubaiyat97@iict.buet.ac.bd

**Table 1. Some pandemics in human history.**

| Pandemics | Number of Deaths (in Millions) |
|---|---|
| Russian Flu | 1 |
| Spanish Flu | 40-50 |
| Asian Flu | 1.1 |
| Third Plague | 12 |
| Acquired immunodeficiency syndrome (AIDS) | 25-35 |
| Swine Flu | 0.2 |
| COVID-19 (27 March 2021) | 2.78 |

In the literature, some systematic studies are reported that focus on AI applied to COVID-19. In the existing studies, the effectiveness of the different attributes of the datasets and the performance of different DL algorithms lack proper explanation. One study [9] reviewed the source, the transmission dynamics, pathogenesis, clinical characteristics of the virus. The study also reviewed the possible ways to overcome the disease in terms of prevention mechanisms as well as treatment options [9]. Another study reviewed the application of big data and AI for managing the effects of COVID-19 [10]. These techniques were used to identify the infected patients, to track the spread of the disease, to the development of drugs and vaccines, to develop diagnosis options and treatment options, *etc.* [10]. A survey paper [11] described the use of AI in fighting against COVID-19. As part of AI, different DL methods, including Generative Adversarial Networks (GAN), Extreme Learning Machine (ELM), and Long/Short Term Memory (LSTM), were described [11]. Different ML and DL algorithms such as random forest, Support Vector Machine (SVM), linear and logistic regression, ensemble eXtreme Gradient Boosting (XG-Boost), and Convolutional Neural Network (CNN) were applied [12]. A survey on DL, deep transfer learning and edge computing was considered [13]. The review work considered the application of DL methods on multi-modal datasets in the context of radiology systems [14]. None of those mentioned above studies simultaneously focus on the spread, detection and diagnosis of COVID-19 using both ML and DL. Moreover, the best ML and DL algorithms are not yet identified for the diagnosis of COVID-19. Furthermore, new studies in ML and DL are going at a very high pace, so there is a need to review the latest findings. Hence, there is a need for a comprehensive survey paper focusing on different COVID-19 datasets and the latest ML and DL algorithms suitable for COVID-19 diagnosis.

This paper provides a systematic and comprehensive review of AI applications in the form of ML and DL in the context of COVID-19. This review article explores the answers to several relevant queries for the successful application of AI on COVID-19.

• The theory and techniques of ML and DL are changing quickly, so the current state of ML and DL needs to be known.

• The capability of ML in solving the forecasting of COVID-19 and in detecting COVID-19 patients need to be understood.

• The capability of DL in diagnosis patients and solving COVID-19 related problems have to be known.

Different learning variants exist under AI. Supervised learning is used to optimise a loss function, considering pre-dicted labels and ground truth, which requires manual annotation. Unsupervised learning is used to find the structure in data considering clustering. Self-supervised learning works with predicted labels and ground truth, which requires a computing process instead of manual annotation. In the case of semi-supervised learning, ground truth requires a mixture of manual annotation and unlabeled data. Transfer learning is used to train from a previously learned model, while multi-task learning is used to optimise simultaneous loss functions without avoiding conflicting gradients caused by individual losses. Weakly-supervised learning requires heuristically labelled data instead of carefully annotated data. On the other hand, multi-modal learning simultaneously deals with multiple data types, including images, texts and electronic health (e-health data). Finally, reinforcement learning takes actions to maximise rewards in a particular situation.

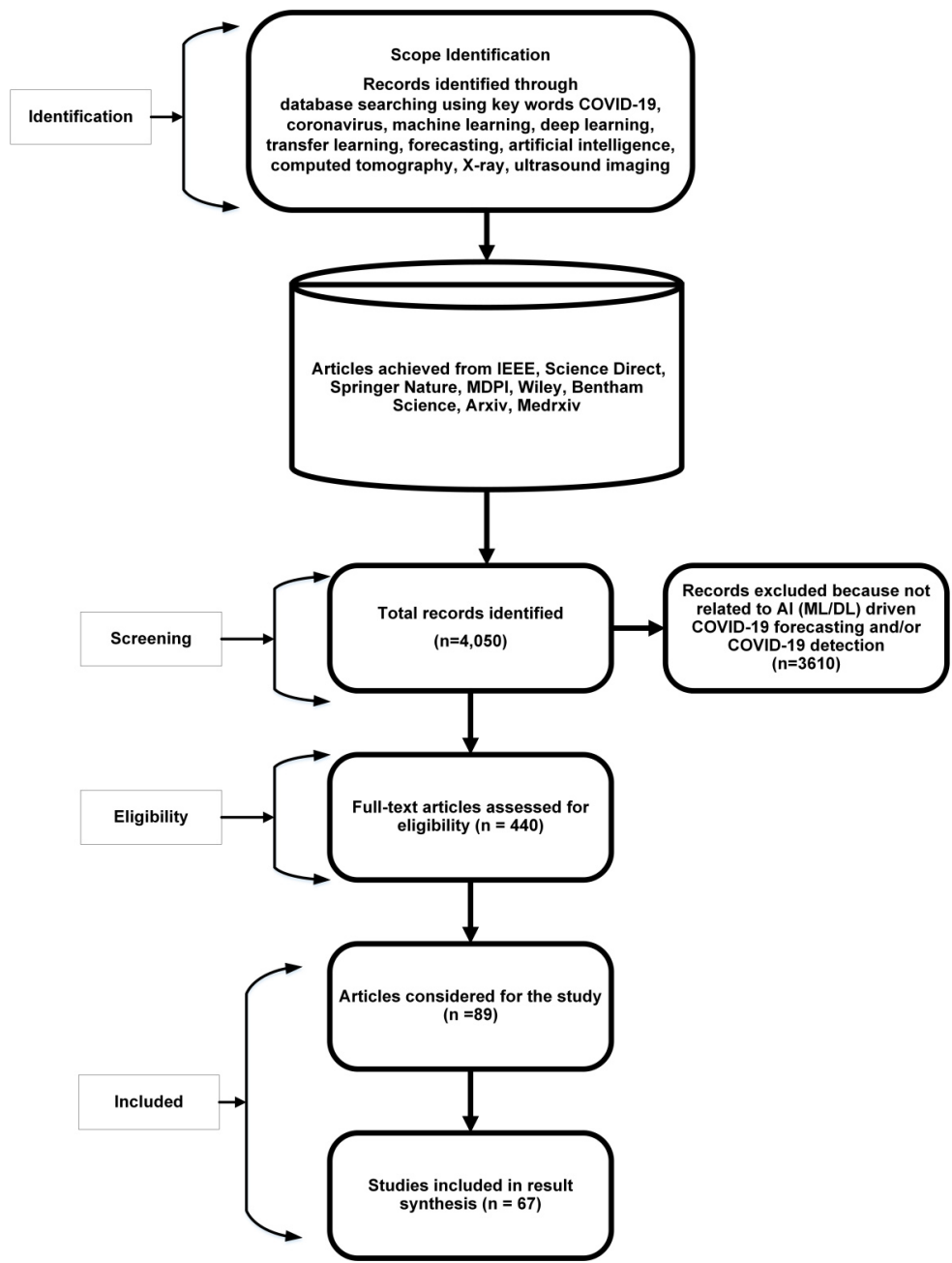

**Fig. (1).** Systematic review through PRISMA guideline.

Fig. (**1**) illustrates the literature survey of this paper following preferred reporting items of systematic reviews and meta-analysis (PRISMA) guidelines. This scoping review was prepared using the PRISMA guidelines as well as the extended version of PRISMA scoping reviews (PRIS-

MA-ScR). Fig. **(1)** shows the flow diagram of the number of records taken into initial consideration, the number of records included and excluded for further consideration with proper justification. The literature search was conducted for all the relevant research papers. One year duration was considered from 1 January 2020 till 27 March 2021. Only English language records were taken into consideration, and papers of other languages were discarded. The records were screened for eligibility according to the Patient, Intervention, Comparator, and Outcome (PICO) criteria. Studies that include all these four criteria were eligible for initial consideration. In this case, patients mean COVID-19 patients, intervention means using AI for screening or diagnosis, comparator means traditional methods, and outcome indicates successful automatic diagnosis. The essential keywords were identified with the use of two Boolean operators, one is OR and the other is AND. The searching was performed for several words, which were COVID-19, coronavirus, artificial intelligence, machine learning, deep learning, forecasting, computed tomography, transfer learning, X-ray, and ultrasound imaging. Special consideration was given to the records related to predicting the spread of COVID-19 using ML regression, classifying suspected patients by ML, and imaging diagnosis by DL. Moreover, the searching was done considering a number of factors of the articles, including the title, abstract and keywords. Papers that focus on basic science, epidemiological and clinical aspects of COVID-19 were not included in this review. Afterward, articles covering duplicate concepts were removed. Next, information extraction and synthesis were performed by extracting data items such as study aim and design, available datasets, categories of AI (ML and DL), and obtained performance results.

A total of 4050 papers on AI on COVID-19 were searched from different reputed publishers and preprints, including IEEE, Elsevier (ScienceDirect), SpringerNature, MDPI, Wiley, Bentham Science, Arxiv, and Medrxiv. The search syntax of this article is shown in the Appendix. Out of these 4050 documents, 3610 articles were excluded manually as those were out of our main research interest. In other words, those did not fulfill the required criteria of the application of AI on COVID-19. A full-text review of the remaining 440 articles was done, and then 89 articles were considered for the study and cited in the paper. Out of these 89 papers, 67 articles are considered for results synthesis as these papers have reported performance metrics, datasets of COVID-19, and performance results of regression or classification algorithms in the context of the use ML or DL on COVID-19. In other words, these 67 papers have reported any of the following issues: ML for forecasting and prediction of COVID-19 cases, features of different modalities for COVID-19 diagnosis, segmentation methods for diagnosis of COVID-19, performance metrics used in AI on COVID-19, and DL for diagnosis of COVID-19 for X-ray/C-T/ultrasound images.

This review work describes the ML regression algorithms used to predict the spread of the disease and the ML classifiers used for classifying COVID-19 patients. Furthermore, the description is provided on the DL methods that are effective for the diagnosis of COVID-19. This paper will facilitate a more in-depth investigation of this subject matter. The rest of this paper is organised as follows. Section 2 provides a discussion on the ML classifiers for classifying COVID-19 patients, as well as ML regressors for predicting confirmed cases, deaths, *etc.*, of COVID-19. Section 3 describes the modalities of DL methods for COVID-19 diagnosis. Section 4 presents the stages of the COVID-19 detection process. Different performance metrics are discussed in Section 5. The comparative performance of different DL algorithms in COVID-19 diagnoses is described in Section 6. The challenges of AI-based solutions are discussed in Section 7. The paper concludes with future directions and concluding remarks.

## 2. APPLICATIONS OF ML FOR COVID-19

This section firstly provides a discussion on machine learning regressors for predicting confirmed cases, deaths, *etc.*, of COVID-19 and then describes the ML classifiers for classifying COVID-19 patients.

### 2.1. ML Regressors for Forecasting COVID-19

This section provides a systematic review of the research works on the application of ML in forecasting the pattern of COVID-19. Table **2** depicts works on ML for forecasting and predicting COVID-19 cases.

The future development trend of this coronavirus has recently gained significant research interest. Several research papers [15-23] focused on using ML techniques on different aspects of COVID-19. The authors used three different mathematical models to perform model-fitting analysis on the confirmed cases of COVID-19 [15]. The models applied in [10] were logistic, Bertalanffy, and Gompertz models. In the analysis of [15], coefficient of determination or R-squared values R2(C) and R2(N) represented the fitting goodness of the cumulative confirmed cases and the new confirmed cases, respectively. Moreover, the above three models were used to predict the number of deaths due to COVID-19. For that, R2 (DC) parameter was used to indicate the fitting goodness of the cumulative number of death cases. Those three models were able to predict the epidemic more accurately in the later stages than in the earlier stages. For Wuhan city of China, the logistic model showed better performance than the other two in predicting the confirmed cases [15]. The work in [17] used a dataset from the repository of Johns Hopkins University. The dataset had three tables, each for confirmed cases, death cases, and recovered cases. Each of the tables had six attributes: province/state, country/region, last update, confirmed cases, death cases, and recovered cases. The study in [17] used a number of ML and DL algorithms such as Support Vector Regression (SVR), Polynomial Regression (PR), Deep Neural Network (DNN), and Long/Short Term Memory (LSTM) for predicting the cases for the next 10 days. The Root Mean Square Error (RMSE) score was used as a metric to evaluate the effectiveness of the algorithms in predicting the COVID-19 cases.

**Table 2. ML for forecasting and prediction COVID-19 cases.**

| Refs. | Adopted Technique | Prediction Result | Objectives | Dataset |
|---|---|---|---|---|
| [15] | Logistic model | In Mainland China, $R^2$(C): 0.9993, $R^2$(N): 0.9183; In Wuhan, $R^2$(C): 0.9991, $R^2$(N): 0.8124. | Prediction of epidemic results of COVID-19. | - |
| | Gompertz model | In Mainland China, $R^2$(C): 0.9934, $R^2$(N): 0.3372; In Wuhan, $R^2$(C): 0.999, $R^2$(N): 0.813. | | - |
| | Bertalanffy model | In Mainland China, $R^2$(C): 0.9993, $R^2$(N): 0.895; In Wuhan, $R^2$(C): 0.9989, $R^2$(N): 0.8105. | | - |
| | Logistic model | In Mainland China, $R^2$(DC): 0.9995; In Wuhan, $R^2$(DC): 0.9993 | Predicting the COVID-19 death toll. | - |
| | Gompertz model | In Mainland China, $R^2$(DC): 0.9997; In Wuhan, $R^2$(DC): 0.9996 | | - |
| | Bertalanffy model | In Mainland China, $R^2$(DC): 0.9991; In Wuhan, $R^2$(DC): 0.9995 | | - |
| [17] | SVR | RMSE for confirmed cases: 27456.47, RMSE for death cases: 1360.47, RMSE for recovered cases: 16762.15 | Prediction of future reachability (next 10 days) of the COVID-2019 across the nations. | [16] |
| [17] | PR | RMSE for confirmed cases: 455.92, RMSE for death cases: 117.94, RMSE for recovered cases: 809.71 | | [16] |
| [18] | ARIMA model of order (1,0,3) | Forecast value of COVID-19 incidence: at 11 February, 2020: 2070.66, at 12 February, 2020: 2418.47. | Evaluate the incidence of new confirmed cases of COVID-2019 in the next 2 days. | [16] |
| [19] | ARIMA model of order (0,1,0) | RMSE (Prediction) for USA: 3963.44, RMSE (Prediction) for Italy: 1258.69, RMSE (Prediction) for China: 59.65 | - | [16] |
| [20] | ARIMA model of order (1,1,0) for confirmed and death cases | $R^2$(Confirmed Case- Italy): 0.761, $R^2$(Death Case- Italy): 0.927 | Forecasting number of daily confirmed cases and deaths. | [21] |
| | ARIMA model of order (2,1,0) for confirmed cases and (0,1,0) for death cases | $R^2$(Confirmed Case- Turkey): 0.817, $R^2$(Death Case- Turkey): 0.714 | | |
| | ARIMA model of order (2,1,1) for confirmed cases and (1,1,2) for death cases | $R^2$(Confirmed Case- Spain): 0.950, $R^2$(Death Case- Spain): 0.958 | | |
| [22] | ARIMA model of order (2,1,2) for Italy, (2,1,0) for China, (1,0,0) for South Korea, (2,3,0) for Iran. | Forecast for 17 days (from 5 March until 21th of March) with 95% confidence interval. | Forecast of confirmed cases of COVID-19 for 17 days. | [16] |
| [23] | ARIMA(1,2,0) for confirmed cases, and ARIMA(3,2,0) for recovered cases | For confirmed cases, ME=17.84, RMSE=514.74, MAE=324.46, MPE=4.26, MAPE=6.25. For recovered cases, ME=80.80, RMSE=186.85, MAE=112.27, MPE=10.57, MAPE=15.60. | | [28] |
| [24] | Logistic Model | Error degrees of freedom: 52 Root Mean Squared Error: 1.9e+03 R-Squared: 0.997, Adjusted R-Squared 0.997 F-statistic *vs.* zero model: 1.61e+04, p-value = 4.02e-77 | Estimated logistic model parameters for China (data up to 11.Mar 2020). | [29] |
| [27] | Hybrid Wavelet ARIMA Model | MAE($\times10^3$): 0.464 (Italy), 0.136 (Spain), 1.341 (USA). RMSE($\times10^3$): 0.630 (Italy), 0.170 (Spain), 1.974 (USA). R-Squared: 0.9985 (Italy), 0.9996 (Spain), 0.9888 (USA). | Prediction of death cases in the next one-month one-month beyond the data sample end date. | [30] |
| | ARIMA Model | MAE($\times10^3$): 1.243 (Italy), 0.693 (Spain), 2.822 (USA). RMSE($\times10^3$): 1.601 (Italy), 0.884 (Spain), 4.103 (USA). R-Squared: 0.9944 (Italy), 0.9089 (Spain), 0.9806 (USA). | | |

In order to model the number of confirmed cases of COVID-19, the authors in [18] applied the autoregressive integrated moving average (ARIMA) model. Using this model, the prediction was made for the next two days. A time series model was used in [19] to predict the COVID-19 outbreak. The length of the time series was 60 days starting from 22 January 2020 to 21 March 2020. The statistical analysis in [19] highlighted the epidemiological stages of different countries so that different policies can be undertaken to combat the epidemic. The study in [20] also used the ARIMA model for modelling the number of confirmed cases and death cases for Italy, Spain, and Turkey for the period 2 February 2020 to 27 March 2020. The authors in [20] reported that the number of confirmed cases in Spain and Italy would decrease in July, while that of Turkey would decrease in September. The study in [22] also used the ARIMA model with Stata version 12 to predict the number of confirmed cases in different countries. The data was considered for the period of 22 January 2020 to 1 March 2020. The authors in [22] reported that Iran and Italy had an unstable trend, while China and Thailand had a stable trend in the number of confirmed cases. A simple automatic forecasting package (AUTOARIMA) of 'R' model was applied in [23]. For this, data of Italian patients were collected in order to predict the number of infected and recovered COVID-19 patients after a lockdown of 2 months. A number of metrics were used, such as Margin of Error (ME), Root Mean Square Error (RMSE), Mean Absolute Error (MAE), Mean Posterior Es-timate (MPE), Median Absolute Prediction Error (MAPE), and Mean Absolute Scaled Error (MASE). The model was used to predict the confirmed cases and an accuracy of 93.75% was achieved, while the recovered cases were predicted with an accuracy of 84.4%. The model highlighted the importance of lockdown and self-isolation in reducing the transmission of the virus. Based on the Italian dataset, the models predicted that lockdown and isolation could decrease the confirmed cases and increase the recovered cases by approximately 35% and 66%, respectively [23].

**Table 3. Some existing ML methods/classifier with its results.**

| Ref. | Method | Validation Method | Data Types | Sample Size | No. of Patients | Results |
|---|---|---|---|---|---|---|
| [31] | Classification (Multilayer perceptron) | Holdout | Clinical | 5644 samples with 111 attributes. The processed dataset has 1091 records and 61 attributes. | 5644 | Accuracy: 93.13%, Recall: 93%, Precision: 93% |
| [32] | Random forest | Cross-validation | Demographics, Clinical | Total 253 samples from 169 patients. Clinical blood test of 49 patients where 24 Covid-19 patients. | 253, 169, 49, 24 | Accuracy: 95.95% Specificity: 96.95% |
| [33] | Support Vector Machine | Holdout | Clinical, Demographics, laboratory features | 336 COVID-19 patients where 26 critical cases. | 336, 220 | Accuracy: 77.5%, AUC: 99%, Specificity: 78.4% |
| [34] | XGBoost classifier | Cross-validation | Blood samples of 75 features, Clinical | 485 samples. | 485 | Accuracy: 90% |

The authors used a logistic model to predict the number of confirmed cases in China and South Korea [24]. According to them, the peak of the COVID-19 epidemic in China was on 8 February 2020, and that of South Korea was on 1 March 2020 [24]. The study also noted that any systematic deviation of a logistic curve indicated the disease was out of control. For example, a deviation of the curve would occur in case there was a start of a second stage of the pandemic [24]. Holt's second-order exponential smoothing method and ARIMA models were used to forecast COVID-19 in India [25]. For this, confirmed cases of 28 Indian states from 30 January 2020 to 21 April 2020 were taken into consideration to fit the models. The models were then used to predict the confirmed and death cases for the next 10 days, starting from 22 April 2020 [25]. A number of models were used to forecast COVID-19 in Brazil [26]. The models adopted were ARIMA, cubist regression (CUBIST), Random Forest (RF), Ridge Regression (RIDGE), and Support Vector Regression (SVR). Moreover, stacking-ensemble learning models with CUBIST, RF, RIDGE, and SVR as base-learners and Gaussian processes as meta-learner were applied. For this, the confirmed cases of 10 Brazilian states from the beginning of the infection up to 18 April 2020 were taken into consideration to fit the models. The models were then used to predict the confirmed and death cases for the next 1, 3, or 6 days starting from 19 April 2020 [26]. A number of metrics, namely Improvement Percentage index (IP), MAE, and Symmetric Mean Absolute Percentage Error (sMAPE) were considered to compare the effectiveness of the models. For the cases considered, SVR and stacking ensemble methods were the most accurate models in forecasting COVID-19 cases in Brazil [26]. The work in [27] modelled the death cases using two different approaches, one is a hybrid model, and another one is the ARIMA model. The dataset considered in [27] had 82 days of observations ranging from 21 January 2020 to 11 April 2020. Five countries, namely the USA, Italy, Spain, UK and France were taken into consideration. Among the 82 days, 66 days were considered as training data and the remaining 16 days were for testing data. The effectiveness of the models was evaluated in terms of mean absolute error (MAE), RMSE and R-squared value.

## 2.2. ML Classifiers for COVID-19

Machine learning (ML) classifiers can be used to classify COVID-19 patients within a dataset. Table 3 shows the list of research work and the associated ML classifiers for COVID-19.

The work used several ML classification algorithms on a dataset of 5644 suspected COVID-19 patients provided by Hospital Israelita Albert Einstein of Brazil [31]. It was shown that Multilayer Perceptron (MLP), XGBoost, and logistic regression successfully classified patients having COVID-19 at a classification accuracy of 91% [31]. In another study, ML classifiers were applied on 253 samples collected from Lanzhou Pulmonary Hospital and Gansu Provincial Hospital, China [32]. A total of 49 attributes from 169 suspected patients were collected for the 253 samples [32]. Among them, 105 samples from 27 patients were COVID-19 positive and were treated as positive samples. Other samples were for lung cancer, common pneumonia, tuberculosis and considered as negative samples. Random forest algorithm achieved an accuracy of 97.95% when applied to 11 features to classify COVID-19 patients from all the samples [32]. A support vector algorithm was applied to a sample of 336 COVID-19 patients to classify critical cases[33]. When the holdout method was used to split the training and testing samples, an accuracy of 77.50% and an area under the receiver operating characteristics curve (AUC) value of 99% were achieved [33]. Another study [34] applied the XGBoost algorithm to the 75 features of 485 samples of COVID-19 patients. When cross-validation was used to split the training and testing samples, an accuracy of 90% was achieved in classifying COVID-19 patients [34].

**Table 4. Distribution of studies for various medical imaging modalities.**

| Image Types | References Studies | No. of Studies |
|---|---|---|
| X-ray | [35, 38-53] | 17 |
| CT | [36, 37, 54-69] | 18 |
| Ultrasound | [70, 71] | 2 |
| Multimodalities | [72, 73] | 2 |

**Table 5. Summary of features, applications, and limitations of different modalities for COVID-19 diagnosis.**

| Imaging Modalities | Important Features and Applications | Limitations | Refs. |
|---|---|---|---|
| X-ray | Data augmentation is used | Not appropriate for multiclass problems. | [50] |
| | DL (neural networks) is considered | Only a limited number of X-ray images are taken into consideration. | [43, 45] |
| | SqueezeNet is used | Not validated for images except for X-ray images. | [35] |
| | Using data augmentation with auxiliary classifier GAN | 1. Small datasets are considered, and image quality is not good. | [48] |
| CT | DL is considered | The method is not validated on a clinical study. | [68] |
| | Using a CAD-based scheme | This is only suitable as an adjuvant method for CT scan images. | [59] |
| | MADE algorithm is used for optimising the attributes of DBM algorithm | MADE-DBM is validated on chest CT datasets. | [36] |
| | A residual learning strategy is used to screen CT images | Not appropriate for detecting patients with early COVID-19 infection. | [62] |
| | Applying Inf-Net to chest CT images. | Accuracy is slightly lower when non-infected regions of the images are taken into consideration. An additional classifier to classify infected and non-infected regions is used to improve the overall accuracy. | [37] |
| Ultrasound | Using DL for analysing LUS images | The dataset has some limitations, for example, not being very large, collected from only a few hospitals in Italy, and the ultrasound was applied to only critically ill patients. There was also poor image quality in some cases. | [70] |
| Multimodalities | COVID-19 diagnosis using both X-ray and CT Images | The combined use of X-ray and CT scan images may not be efficient in real-time practical applications. | [72] |
| | Combination of ML, DL and data augmentation for COVID-19 detection | Feasible only for limited datasets. | [73] |

## 3. MODALITIES OF DL FOR DIAGNOSIS OF COVID-19

Medical images are scarce for the diagnosis of COVID-19. Still, studies on the application of DL methods to the X-ray and CT scan images to diagnose COVID-19 patients. X-ray radiations can move past different objects along with the human body. Hence, X-rays are useful in taking images of the inside structures of the human body. A CT scan can generate cross-sectional images of the body using computers and a form of rotating X-ray machinery. CT images provide more information than a corresponding X-ray image of the same object. Ultrasound images are generated as real-time videos with the use of sound waves. Besides all these, some studies apply multiple image modalities to detect COVID-19 patients. Table **4** shows the number of research works on different modalities for the detection of COVID-19. It can be seen that most research papers are on X-ray and CT images, and only a few are on ultrasound and mix modalities.

Table **5** presents different research works on the application of DL on different images. The major features, the applications, and the limitations of the studies are also presented

in Table **5**. It can be seen from Table **5** that deep Bayes-SqueezNet was applied to X-ray images [35]. In another study, Memetic Adaptive Differential Evolution (MADE) algorithm was introduced. This MADE algorithm is for the optimisation of the parameters of Deep Boltzmann Machine (DBM) model [36]. Lung infection segmentation deep network (Inf-Net) was used in [37] in order to find the infected areas of CT images.

## 4. STAGES OF COVID-19 DIAGNOSIS

This section describes the process of data-driven COVID-19 diagnosis. Fig. (**2**) illustrates a framework for COVID-19 diagnosis and that there are five major layers of operation. Firstly, the input database is studied, and then a suitable imaging technique is selected by physicians or by the telemedicine system. Next imaging-based techniques: X-ray, CT scan, or ultrasound imaging are used. AI-based processing and optimisation are then performed. Finally, results are estimated and reported. The overall diagnosis has some other process steps. These steps are preprocessing, segmentation, feature selection, classification, and detection. The following literature review is made on preprocessing and segmentation stages.

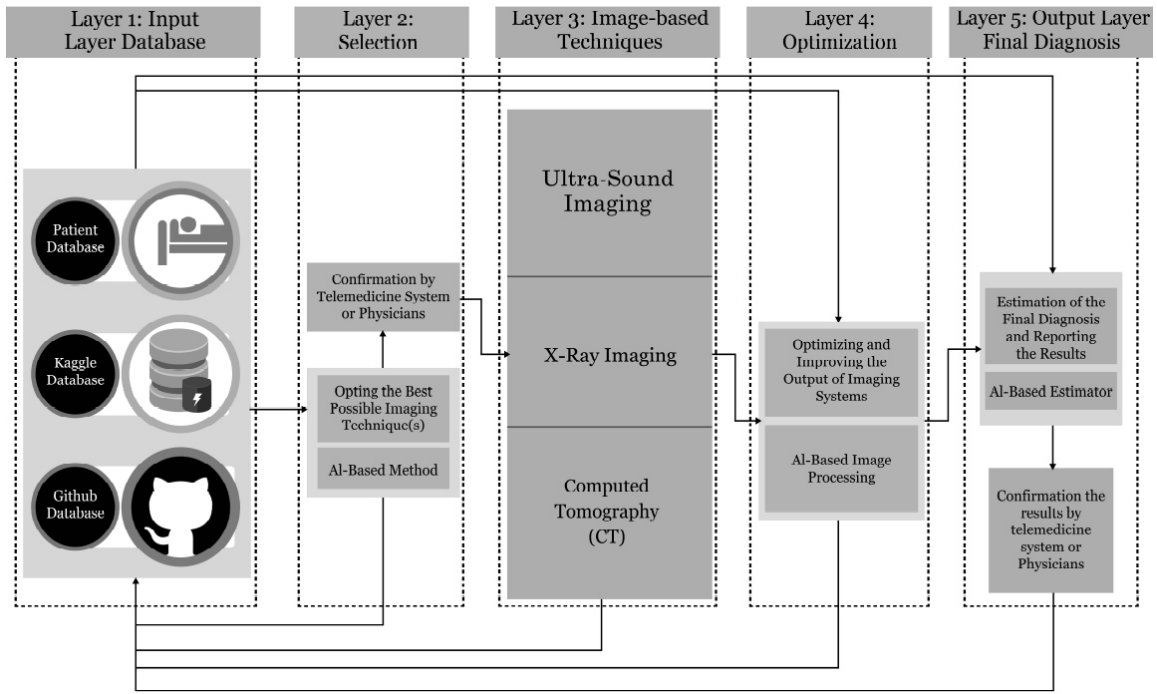

**Fig. (2).** Framework for COVID-19 diagnosis. (*A higher resolution / colour version of this figure is available in the electronic copy of the article*).

Firstly, the preprocessing steps are considered. 2D UNet was used for the preprocessing of CT images in [63]. Online data augmentation was used in order to overcome the overfitting problem caused by a limited dataset of CT images. Two data augmentation techniques were random affine transformation and colour jittering [54]. Data augmentation was also done for a composite model of GAN called CovidGAN [48]. In that case, CovidGAN was applied for the synthesis of COVID-19 and normal CXR classes. Other forms of data augmentation methods known as rotation and translation are reported [49]. Data augmentation using facile image transformation was described in a study [73]. The facile transformation was in the form of scaling, rotation at various angles, and flipping operations [73]. Image ROI was extracted using Visual Basic .NET (VB-Net) for the case of 618 CT images [74]. Next, CNN was used to classify COVID-19 and Influenza-A viral pneumonia and normal people. A threshold approach was used to extract ROI of images, and then a modified inception network algorithm was applied to classify COVID-19 and viral pneumonia [75]. VB-Net was used to extract Region of Interest (ROI) of images on which a random forest classifier was applied to classify COVID-19 and common pneumonia [76].

Next, image segmentation is considered. In the diagnosis of COVID-19, segmentation was used in a number of research papers. Several segmentation methods based on DCNN were used [77-82]. A fully CNN (FCN) was reported for semantic segmentation [77], while some studies focus on a number of variants of FCN [78-81] for segmentation. A study [83] modified V-Net and used 3D U-Net [84] for segmentation. It can be noted that U-Net was developed based on CNN and it was adapted to achieve better segmentation in the context of COVID-19 diagnosis. Some studies reported the automatic segmentation of COVID-19 pneumonia lesions from images [85]. Another study [57] reported the use of U-Net [78] for segmenting lungs from CT scans to classify COVID-19 pneumonia and traditional pneumonia. The work in [86] and in [87] applied U-Net algorithm for segmentation. Another study used UNet++ [88]. This UNet++ was also used for detection [56] and segmentation [55]. In [89], A VB-Net was formed using V-Net [84] and bottleneck [90] for segmentation [89]. However, none of the above segmentation studies considered noise in the annotations.

Table **6** summarises different segmentation methods used in the diagnosis of COVID-19. It can be seen from Table **6** that the popular segmentation methods used in the literature are classic U-Net [54, 57, 69, 86, 87, 91], UNet++ [55, 56], VB-Net [89]. X-rays are less expensive and easier to obtain than CT images. However, the segmentation of X-ray images is more challenging than CT images. This is because, in X-ray 2D images, the ribs are projected on soft tissue, confounding image contrast. Hence, there is no method for segmenting X-ray images in the context of COVID-19 diagnosis. A study employing Attenuation-U-Net to segment lung image regions for detecting pneumonia may be applicable for detecting COVID-19 as well [94].

**Table 6. Summary of segmentation methods for COVID-19 diagnosis.**

| Modality | Target ROI | Method | Remarks | Refs. |
|---|---|---|---|---|
| CT | Lung | U-Net | The method is weakly supervised with pseudo labels | [54] |
| | Lesion | UNet++ | - | [55] |
| | Lesion and Lung | UNet++ | Joint classification and segmentation | [56] |
| | Lesion | U-Net | - | [57] |
| | Lesion and Lung lobes | U-Net | - | [69] |
| | Lesion and Lung | U-Net | - | [86] |
| | Lung, Lesion and Lung lobes | U-Net | - | [87] |
| | Lung, Lesion, Lung lobes and Lung segments | VB-Net | Human-in-the-loop | [89] |
| | Lesion and lung | Commercial Software/U-Net | Combination of 3D and 2D methods | [91] |
| | Lesion, Lung Bronchus and Trachea | Commercial Software | - | [92] |
| | Lesion | Threshold-based region growing | - | [93] |

**Table 7. Performance metrics used in each selected primary study.**

| Performance Matrix | Refs. | Number of Studies |
|---|---|---|
| Accuracy | [35, 36, 38, 41, 43-55, 58-60, 62-65, 67, 68, 70, 72, 99, 100] | 30 |
| Recall | [35-38, 41, 43-53, 55, 57-60, 62, 64, 65, 67, 68, 70, 72, 73, 99, 100] | 31 |
| Specificity | [35, 37, 38, 43-46, 48-53, 55, 57, 59, 62, 64, 67, 68, 73, 99, 100] | 23 |
| Precision | [35, 36, 38, 41, 44-51, 54, 55, 58-60, 62-65, 68, 70, 72, 73, 99] | 26 |
| Negative predictive value (NPV) | [54, 55, 59, 63, 68, 73] | 6 |
| F-measure | [35, 36, 38, 41, 43-51, 58, 60, 62, 65, 67, 68, 70, 72, 73, 99] | 23 |
| AUC | [36, 44, 47, 50, 51, 57-60, 62-65, 67, 100] | 15 |
| Matthews correlation coefficient (MCC) | [35, 47, 50, 73] | 4 |
| Mean absolute error (MAE) | [37] | 1 |

## 5. PERFORMANCE METRICS

A number of metrics were used to classify and diagnose COVID-19. Table **7** shows a list of performance metrics used in the literature of COVID-10 diagnosis. Some of these metrics are accuracy, recall, precision, specificity, negative value predictor, F-measure, *etc.* [95-98]. These are expressed below. Accuracy indicates the correctness of classifying normal cases as normal and abnormal cases as abnormal. Recall or sensitivity refers to the ratio of the number of correctly classified COVID-19 positive patients to the total number of suspected patients. Specificity refers to the accuracy of the prediction of negative or normal cases. Precision or positive predictive value is the ratio of accurately classified positive cases to the total predicted positive cases. The negative predictive value is the ratio of the correctly classified negative samples to the total predicted negative samples. F-measure is the harmonic means of precision and recall. The AUC is a measure of how well classification is performed between positive and negative cases. Mean Absolute Error (MAE) is calculated as the average of the absolute difference of the predictive and actual sample values.

## 6. COMPARATIVE ANALYSIS OF DL FOR COVID-19

This section provides a systematic review of the research works on the application of DL in the diagnosis of COVID-19. For this, both X-ray and CT scan images are taken into consideration.

A number of research papers [38-42, 54-59, 101] focused on the use of DL techniques in the diagnosis of COVID-19 patients. Table **8** shows different DL methods for the diagnosis of COVID-19 using X-ray images. The study in [38] used the Kaggle's Chest X-ray dataset [102] and the dataset in [103]. For the experiments, X-ray images of 50 patients (abnormal) and 50 normal persons were taken into consideration. The study [38] reported that the best classification accuracy of 98.0% was obtained using the residual neural network 50 (ResNet50) algorithm, while InceptionV3 and Inception-ResNetV2 achieved classification accuracy values of 97% and 87%, respectively. Wang *et al.* [39] applied a deep convolutional neural network-based model to X-ray images of suspected COVID-19 patients and achieved a testing accuracy of 83.50%. The two online datasets [104] used had a total of 5941 chest X-ray images collected from 45 COVID-19 patients, 660 people with viral pneumonia, 931 people having bacterial pneumonia, and 1203 healthy people. Zhang *et al.* [42] applied ResNet based model to X-ray images. The ResNet model was applied for two purposes [42]. The first purpose was to perform the classification of suspected patients into normal (disease negative) and abnormal (disease positive) cases. The second purpose was to detect anomaly in classifying suspected patients. For data samples of 70 patients and 1008 normal (non-patients), the obtained sensitivity, specificity, and AUC values were 96%, 70.7%, and 95.2%, respectively [42]. From the above discussion, it can be seen that the highest accuracy is obtained by

**Table 8. DL for diagnosis of COVID-19 for X-ray images.**

| Refs. | Dataset | Methods | Accuracy | Recall |
|---|---|---|---|---|
| [38] | [105, 106] | ResNet 50 | 98% | 96% |
| | | Inception V3 | 97% | 94% |
| | | Inception ResNetV2 | 87% | 84% |
| [39] | [103, 107, 108] | COVID-Net network architecture | 92.6% | 87.1% |
| [40] | [103] | Bayesian ResNet50V2 | 92.9% | - |
| [41] | [103, 105] | AlexNet for 4 classes | 66.67% | 66.67% |
| | | AlexNet for 3 classes | 85.19% | 85.19% |
| | | AlexNet for 2 classes | 100% | 85.19% |
| | | GoogleNet for 4 classes | 80.56% | 80.56% |
| | | GoogleNet for 3 classes | 81.48% | 81.48% |
| | | GoogleNet for 2 classes | 100% | 100% |
| | | ResNet18 for 4 classes | 69.46% | 69.46% |
| | | ResNet18 for 3 classes | 81.48% | 81.48% |
| | | ResNet18 for 2 classes | 100% | 100% |
| [42] | Collected from 6 institutions | Deep anomaly detection model | - | 96% |
| [43] | [103, 109] | DarkCovidNet (multi-class classification task) | 87.02% | 85.35% |
| | | DarkCovidNet (binary classification task) | 98.08% | 95.13% |
| [44] | [110], Custom | CovXNet (COVID-19/ Normal) | 97.4% | 97.8% |
| | | Residual Network (COVID-19/ Normal) | 92.1% | 93.4% |
| | | CovXNet (COVID-19/ Bacterial Pneumonia) | 94.7% | 94.4% |
| [45] | [105, 106] | CoroNet | 89.6% | - |
| [99] | [105, 111] | MobileNetV2 | 98.54% | 97.87% |
| | | SqueezeNet | 97.81% | 100% |
| | | SqueezeNet & MobileNetV2 | 99.27% | 100% |
| [35] | [103, 110] | Bayes-SqueezeNet | 98.3% | - |
| [48] | [105, 107, 112] | Actual data (CNN-AD) | 85% | 65% |
| | | Actual data & Synthetic augmentation (CNN-SA) | 95% | 90% |
| [49] | [102, 105, 109, 113-115] | DenseNet201 | 99.70% | 99.70% |

ResNet18 being 100%, followed by dense convolutional network (DenseNet)-169 being 99.70%.

Table **9** shows the application of DL methods to CT images in order to diagnose COVID-19. The work in [55] applied UNet++ based segmentation model to chest CT scan images of a dataset with 51 patients and 55 normal people. This model was used to segment COVID-19 related lesions which prepared the final label of COVID-19 patients or non-patients. The model obtained accuracy, sensitivity, and specificity of 95.2%, 100%, and 93.6%, respectively. This model was also applied to another dataset of 16 viral cases of pneumonia and 11 normal people and successfully made all the classifications [55]. A hybrid UNet++ and ResNet50 classification model was applied to a dataset containing CT scan images of 1136 cases where 723 were COVID-19 patients [56]. A ResNet50 classification model was applied to a dataset containing CT scan images of 4356 cases where 1296 were COVID-19 patients, 1735 were community-acquired pneumonia patients, and 1325 non-pneumonia patients [57]. In this study [57], ResNet50 was used to classify 2D slices with shared weights. The model achieved a sensitivity, specificity, and AUC value of 90%, 96%, and 96%, respectively. A DL approach using ResNet50 termed as DeepPneumonia was applied to a dataset of 88 COVID-19 patients, 101 patients with bacterial pneumonia, and 86 normal people [58]. In this study [58], CT scan images were used to generate slices of complete lungs which were fed to the input of DeepPneumonia algorithm. The model achieved an accuracy of 86% in classifying COVID-19 and bacterial pneumonia, and an accuracy of 94% in classifying COVID-19 patients and healthy people. For the case of DenseNet169, an accuracy value and a recall value of 99.80% was achieved [60]. A model termed CovXNET was used to extract numerous attributes from X-ray images [44]. For this CovXNET was ap-

**Table 9. DL for diagnosis of COVID-19 for CT images.**

| Refs. | Dataset | Methods | Accuracy | Recall |
|---|---|---|---|---|
| [54] | Custom | 2D UNet and DeCoVNet | 90.8% | 90.7% |
| [55] | Dataset from Renmin Hospital of Wuhan University | UNet++ | 95.2% | 100% |
| [56] | - | UNet and CNN | - | 97.4% |
| [57] | [116] | ResNet 50 | - | 90% |
| [58] | [91] | ResNet 50 | 86% | - |
| [59] | CovNet[117, 118] | AlexNet | Training: 82.60% Validation: 78.92% | Training: 93.38%, Validation: 89.21% |
| | | ResNet 50 | Training: 98.28% Validation: 94.12% | Training: 96.56% Validation: 90.20% |
| [60] | [119, 120] | VGG16 | 99.20% | 99.30% |
| | | VGG19 | 98.85% | 99.50% |
| | | Xception | 99.60% | 99.80% |
| | | ResNet 50 | 99.20% | 99.60% |
| | | ResNet 50V2 | 99.35% | 99.30% |
| | | DenseNet121 | 99.45% | 99.60% |
| | | Inception V3 | 99.60% | 99.60% |
| | | Inception ResNet V2 | 99.65% | 99.70% |
| | | DenseNet 169 | 99.80% | 99.80% |
| [72] | [102, 103, 108] | VGG16 | 91% | 94% |
| | | VGG19 | 90% | 94% |
| | | Xception | 96% | 98% |
| | | ResNet 50 | 98% | 96% |
| | | ResNet 50V2 | 96% | 96% |
| | | DenseNet121 | 99% | 98% |
| | | Inception V3 | 95% | 99% |
| | | Inception ResNet V2 | 94% | 96% |
| [64] | Custom | weakly supervised deep learning | 96.2% | 94.5% |
| [65] | From hospitals in Shandong Province | AD3D-MIL | 94.3±0.7% | 90.5±0.5% |
| [36] | Custom | MADE-DBM (For multilevel classification) | 96.19% | 96.23% |

plied on two datasets and exhibited better performance than other models. One dataset collected at Guangzhou Medical Center of China had 5856 X-ray images, where 1583 were for normal people without any infection, 1493 for traditional pneumonia, and 2780 were for bacterial pneumonia [110]. The second dataset collected at Sylhet Medical College of Bangladesh had 305 X-ray images. In mobile devices, the MobileNetV2 model was used [99]. An accuracy of 100% and 99.27% were obtained in classifying COVID-19 patients and pneumonia patients, respectively [99]. Transfer learning, image augmentation, and CNN were applied on a dataset combing existing datasets and a self-generated dataset by the authors [49]. In the dataset, the instances of COVID-19 patients, viral pneumonia patients, and normal people were 423, 1845, and 1579, respectively [99]. The accuracy values were 99.7% and 97.9% when used to classify COVID-19, and classify 3-class cases (normal, viral pneumonia, and COVID-19), respectively [49].

From the above discussion, it can be seen that the highest accuracy was obtained by DenseNet169 being 99.80% followed by Inception ResNet v2 being 99.65%. In order to summarise the findings, Table **10** presents the best results found in Tables **6**, **7**, **8** and **9**. Apart from the best models for X-ray and CT imaging, Table **10** shows the segmentation methods used and the performance metrics applied in those studies. From Table **10**, it can be seen that the study [41] achieves the best accuracy, precision, recall and F1 score values of 100% for multiclass classification using X-ray images with GAN segmentation. On the other hand, the study in [60] achieves the best accuracy, precision, recall and F1 score values of 99.80% for binary classification using CT images with BConvLSTM, U-Net and GAN segmentations. Fig. **(3)** illustrates the accuracy values of different DL classifiers when applied to X-ray and CT scan images. It can be seen that the highest accuracy value for X-ray images is 100% and that for CT images is 99.80%.

## 7. CHALLENGES OF AI-BASED SOLUTIONS

In this section, the challenges that exist in using ML and DL to predict, detect and manage COVID-19 are discussed.

During the COVID-19 pandemic, several steps were taken by different countries to fight against the virus spread.

**Table 10.** Overall summary of the DL in COVID-19 diagnosis.

| Refs. | Images of Used Dataset | Imaging Modalities | Segmentation | Adopted Model | Precision | Recall | F1 Score | Accuracy |
|---|---|---|---|---|---|---|---|---|
| [41] | COVID-19: 69, Normal: 79, Bacterial Pneumonia: 79, Viral Pneumonia: 79 | X-ray | GAN | ResNet 18 | 100 | 100 | 100 | 100% |
| [60] | COVID-19: 1232, Healthy: 1668 | CT | BConvLSTM U-Net and GAN | DenseNet 169 | 99.80 | 99.80 | 99.80 | 99.80% |
| [49] | COVID-19: 423, Normal: 1579, Viral Pneumonia: 1485 | X-ray | - | DenseNet 201 | 99.70 | 99.70 | 99.70 | 99.70% |
| [60] | COVID-19: 1232, Healthy (Normal): 1668 | CT | BConvLSTM U-Net and GAN | Inception ResNet V2 | 99.60 | 99.70 | 99.65 | 99.65% |
| [60] | COVID-19: 1232, Healthy (Normal): 1668 | CT | BConvLSTM U-Net and GAN | Inception V3 | 99.60 | 99.60 | 99.60 | 99.60% |
| [60] | COVID-19: 1232, Healthy (Normal): 1668 | CT | BConvLSTM U-Net and GAN | Xception | 99.40 | 99.80 | 99.60 | 99.60% |
| [99] | COVID-19: 295, Normal: 65, Pneumonia: 98 | X-ray | Fuzzy Color and Stacking | SqueezeNet & MobileNetV2 | 98.89 | 98.33 | 98.58 | 99.27% |

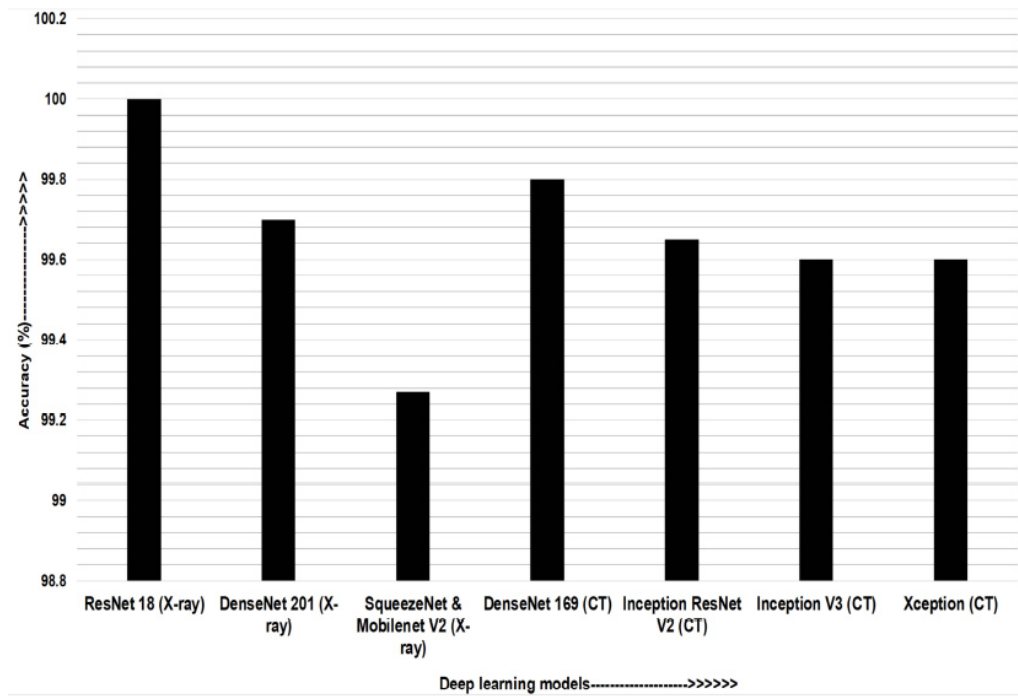

**Fig. (3).** Bar diagram of the accuracy values of different classifiers for X-ray and CT images. (*A higher resolution / colour version of this figure is available in the electronic copy of the article*).

Some of these were lockdown and social distancing. Forming regulations was vital in providing guidelines for citizens, business people, medical staff, scientists and others to prevent the virus's spread. Although ML and DL methods have great potential in managing COVID-19, the effectiveness of the methods depends on the datasets. Some datasets contain incomplete and erroneous labels, which creates a challenge to the disease prediction. Manually labelling the X-ray or CT images can be time-consuming as well as costly. Hence, self-supervised DL and transfer learning algorithms can be useful. Most of the AI-based DL methods require large-sized reliable data, particularly data for training the models. Analysing the training data samples is a time-consuming operation requiring high-performance computing devices. Trained medical experts are also required for labelling data samples. Despite the requirement of such datasets, there is a scarcity of large reliable and balanced datasets on COVID-19. Sometimes excess data can make extracting rele-vant information difficult. Finding the actual treatment results of suspected patients is also a difficulty. Real-life medical images usually have various types of noise and distortions, which require different types of image denoising. The presence of this wrong information hinders the effectiveness of the performance of AI methods. Moreover, there is no benchmark dataset for evaluating the performance of different algorithms. AI needs to be used appropriately in identifying and separating erroneous information from fact. This is particularly useful in the context of COVID-19, where there are many false or unverified information, reports, audio files, video files, social media statuses, blogs and others. For practical application, ML or DL-based COVID-19 diagnosis system needs to be integrated into the X-ray or CT scan system, which will ensure that good quality images are acquired at low radiation, making the overall system efficiency. In order to get the effectiveness of AI in fighting COVID-19, there is a need for the intersection of information and communication (ICT) or computer science specialists and virology or medical science experts. Developing a robust collaborative platform of experts from different disciplines is still a challenging task. Another challenge is the data protection issue which requires preserving human rights and privacy.

This paper reviews the latest applications of AI-based methods in forecasting the spread, classifying and diagnosing of COVID-19. The findings of this review depend on the information provided in the literature which is varying at a rapid pace. This review does not focus on the epidemiological and clinical aspects of COVID-19. Moreover, focusing on COVID-19 treatment and vaccine development is out of this current review's scope and left as future work.

## CONCLUSION

This paper reviews the application of ML and DL methods in fighting against COVID-19. The systematic review is performed following PRISMA guidelines. The review reveals that ARIMA models of different orders, PR, RIDGE, SVR, logistic models, and the hybrid wavelet ARIMA model effectively predict the number of confirmed cases. It is further found that multilayer perceptron, SVM, random forest, XGBoost, *etc.* are useful in classifying COVID-19 patients from normal people. In one particular case, random forest achieves a classification accuracy of 95.95%. In DL-based COVID-19 diagnosis, there are several stages, including image preprocessing, segmentation, feature extraction and classification. This review work also shows that DL methods such as CNN, ResNet, COVIDNet, VGG-16, VGG-19 and hybrid neural networks can successfully diagnose COVID-19 when used on X-ray and CT scan images. For the case of X-rays images, ResNet 18, GoogleNet and AlexNet algorithms exhibit an accuracy of 100% when applied to a small dataset of 148 samples and when binary classification is done to identify COVID-19 patients and normal people. However, when classification is done for normal, COVID-19, and viral pneumonia cases on a dataset of 3,487 samples, maximum accuracy of 99.70% is achieved using DenseNet 201. For the case of CT images, DenseNet 169 shows a binary classification accuracy of 99.80% when applied to a dataset of 2,900 image samples. On the other hand, an accuracy of 97% is achieved for the case where U-Net++ is applied to 58,924 ultrasound images.

In future, a number of challenges have to be overcome to utilise the benefits of AI practically. Some of the important challenges are regulations, lack of reliable large datasets, incorrect data, noisy data, lack in the intersection of AI and medical fields, and data privacy. Furthermore, high-quality medical datasets with large sample sizes will be vital in near future. Moreover, the datasets should include COVID-19 patients of different stages, where the presence of borderline patients will be important to evaluate the effectiveness of a classifier. The reliability of COVID-19 diagnosis can be improved by using multiple imaging techniques, for example, ultrasound, X-ray, CT scan, and Magnetic Resonance Imaging (MRI) for each suspected patient. This is because such a multi-modal system takes advantage of the important features of each imaging technique. There are a lot of COVID-19 medical images without having appropriate labels, therefore applying unsupervised learning algorithms can be useful. Finally, it is expected that with the advent of novel ML and DL algorithms, the world will be able to manage COVID-19 pandemic successfully.

## APPENDIX

In Appendix, the search criteria for this review are discussed. Initially, 4050 research papers were selected for the duration of 1 January 2020 to 27 March 2021 from Google Scholar based on the following criteria:

(("COVID-19") OR ("coronavirus"))AND (("machine learning") OR ("deep learning") OR ("transfer learning")OR("forecasting"))AND("artificial intelligence") AND(("computed tomography") OR ("X-ray") OR ("ultrasound imaging")).

## CONSENT FOR PUBLICATION

Not applicable.

## STANDARDS OF REPORTING

PRISMA guidelines were followed.

## FUNDING

None.

## CONFLICT OF INTEREST

The authors declare no conflict of interest, financial or otherwise.

## ACKNOWLEDGEMENTS

The work has been carried out at the Institute of Information and Communication Technology (IICT) of Bangladesh University of Engineering and Technology (BUET), Bangladesh. Hence, the authors thank BUET for providing research facility.